\documentclass[aps,nofootinbib]{revtex4-2}
\usepackage{amssymb}
\usepackage{eurosym}
\usepackage{amsfonts}
\usepackage{geometry}
\usepackage{color}
\usepackage[normalem]{ulem}
\usepackage{bbm}
\usepackage{soul}
\usepackage{color}
\usepackage{mathtools}

\def\be{\begin{equation}}
\def\ee{\end{equation}}
\def\bea{\begin{eqnarray}}
\def\eea{\end{eqnarray}}
\begin{document}
\title{From Snyder space-times to doubly $\kappa$-dependent Yang quantum phase spaces and their generalizations}
\author{Jerzy Lukierski$^a$, Stjepan Meljanac$^b$, Salvatore Mignemi$^c$, Anna Pacho\l$^d$ and Mariusz Woronowicz$^a$}
\affiliation{
$^a$ Institute of Theoretical Physics, Wroc\l aw University, pl.~Maxa Borna 9, 50-205 Wroc\l aw, Poland\\
$^b$ Division of Theoretical Physics, Rudj{}er
Bo\v{s}kovi\'c Institute, Bijeni\v{c}ka~c.54, HR-10002~Zagreb, Croatia\\
$^c$ Dipartimento di Matematica, Universit\`a di
Cagliari via Ospedale 72, 09124 Cagliari, Italy and INFN, Sezione di
Cagliari, 09042 Monserrato, Italy\\
$^d$ Department of Microsystems, University of South-Eastern Norway, Campus Vestfold, Norway}
\begin{abstract}
We propose the doubly $\kappa$-dependent  Yang quantum phase space which describes the generalization of $D = 4$ Yang model. We postulate that such model is covariant under the generalized Born map, what permits to derive this new model from the earlier proposed $\kappa$-Snyder model. Our model of $D=4$ relativistic Yang quantum phase space depends on five deformation parameters which form two Born map-related dimensionful pairs: $(M,R)$ specifying the standard Yang model and $(\kappa,\tilde{\kappa})$ characterizing the Born-dual $\kappa$-dependence of quantum space-time and quantum fourmomenta sectors; fifth parameter $\rho$ is dimensionless and Born-selfdual. In the last section, we propose the Kaluza-Klein generalization of $D=4$ Yang model {and the new quantum Yang models described algebraically by quantum-deformed $\hat{o}(1,5)$ algebras.}
\end{abstract}
\maketitle
\section{Introduction}
Quantum Gravity (QG) describing the quantum dynamics of curved space-times interacting with matter sources leads to the introduction of noncommutative (NC) space-time coordinates $\hat{x}_\mu$ ($\mu=0,\ldots,3$) (see e.g. \cite{DFR, Majid2009}) and the NC relativistic quantum phase spaces $(\hat{x}_\mu,\hat{q}_\mu)$ (see e.g. \cite{Low, Lukierski2021PoS, AP_EPJC2023}).
Already in 1947 Snyder \cite{Snyder} and Yang \cite{Yang} proposed the first examples of Lorentz-covariant NC space-times and NC relativistic quantum phase spaces.
In the passage from classical to quantum gravity the algebraic QG models describing NC geometry should contain the explicit dependence on the Planck constant $\hbar$ (see e.g. \cite{Bronstein, Chamseddine, Aschieri}). We add that in the limit $\hbar\rightarrow 0$ one gets from QG the commutative classical models (see e.g. \cite{Aschieri,Drinfeld, Connes, Beggs, Arzano}). 

Snyder model \cite{Snyder}, which can be described by $\hat{o}(1,4)$ algebraic relations, introduces the  NC model of quantum relativistic space-time. Such algebraic model does not provide a dynamical input determining e.g. the time evolution of the system. However, there were also introduced  quantum dynamical particle models providing the Snyder algebra \cite{Snyder} derived from the constraints obtained from the quantized Dirac brackets (see e.g. \cite{17a,17b}).\\
The class of relativistic quantum phase spaces with NC quantum space-time coordinates $\hat{x}_\mu$ and commuting fourmomenta $q_\mu$ was proposed already in \cite{Snyder}. Further it has been shown (see e.g. \cite{5a,5aa}) that in such class of Snyder phase spaces there is a freedom in  commutation relations which can be described by an arbitrary function $F(q^2)$ \cite{5a}. 

The D=4 Yang model was introduced in \cite{Yang} and incorporates in one quantum relativistic phase space algebra the NC quantum space-time coordinates $\hat{x}_\mu$ as well as curved NC quantum fourmomenta $\hat{q}_\mu$. This quantum phase space depends on the pair of dimensionful parameters M $([M]=L^{-1})$\footnote{In this paper we systematically put $c=1$.} and R $([R]=L^1)$, describing the curvatures of NC space-time and NC fourmomenta space which are both related by the Born map \cite{Born1}, \cite{Born2}.

Further new versions of Snyder model were obtained by choosing various realizations of Snyder algebra. From the algebraic point of view these new models remain the same, however due to the physical interpretations of the parameters one obtains effectively new physical models in D=4 dimensions. In particular, one can embed into one algebraic model two types of quantum space-time noncommutativity: the one characterizing the Snyder model and the other which provides the $\kappa$-deformation of Minkowski space-time coordinates\footnote{The NC $\kappa$-deformation of Minkowski space-time can be rigorously derived from quantum deformation of Poincare algebra and its Hopf-dual quantum Poincare group.} \cite{1991pn}-\cite{LRZkappa} with the parameter $\kappa$ often identified with the Planck mass (see e.g. \cite{Planckmass}). 
{Consistent description of the quantum space-times with their noncommutativity described by the sum of Snyder and $\kappa$-Minkowski terms\footnote{Such models provide the so-called $\kappa$-Snyder space-times.} have been proposed earlier, see e.g. \cite{kappa-Snyder1}-\cite{BP_SIGMA2014}. Moreover, the introduction of supplementary $\kappa$-deformation terms in Snyder models has been recently explained \cite{MM_PLB814}-\cite{LMMP_PLB838} as following from the suitable modification of standard $\hat{o}(1,4)$ realizations.} By using two isomorphic realizations of $\hat o(1, 4)$ algebra one obtains two different $D=4$ physical models (Snyder and $\kappa$-Snyder).
If the linear change of quantum space-time coordinates contains the Lorentz symmetry generators (for details and explicit transformation formulae see \cite{MM_PLB814}) it  describes the passage from Snyder to $\kappa$-Snyder models.

In the present paper (see also \cite{lukpa}, \cite{MMM2024}) we added the $\kappa$-Minkowski terms to the Yang model, what provides new Lorentz-covariant quantum relativistic phase spaces described by doubly $\kappa$-dependent Yang models\footnote{The name "doubly $\kappa$-deformed Yang models" (see e.g. \cite{MMM2024}) we abandoned in this paper because it could suggest that new models require the deformation of Lie-algebraic structure $\hat{o}(1,5)$ which describes standard Yang model.}. 
Since in Yang models we can introduce the pair of different $\kappa$-Minkowski terms, in NC space-time and fourmomenta sectors, we need two independent mass-like parameters $\kappa$ and $\tilde{\kappa}$.
We add that for simplicity, instead of doubly $\kappa$-dependent Yang models, we will use the shorter notation $(\kappa,\tilde{\kappa})$-Yang models. 
It appears that such pair of $\kappa$-dependence is related by the generalized Born map $\tilde{B}$, acting as follows (see also \cite{lukpa}, \cite{MMM2024})
\begin{equation}
\hat{x}_\mu\rightarrow\hat{q}_\mu,\qquad \hat{q}_\mu\rightarrow-\hat{x}_\mu,\qquad M\leftrightarrow R,\qquad \kappa\leftrightarrow\frac{1}{\tilde{\kappa}}.
\label{bm1}
\end{equation}
Further, by using Jacobi identities one can show that in $(\kappa,\tilde{\kappa})$-Yang models one can still introduce (besides $M$, $R$, $\kappa$ and $\tilde{\kappa}$) one additional fifth dimensionless parameter $\rho$ $([\rho]=L^0)$, which can be linked with the so-called TSR ("Triply Special Relativity") model \cite{TSR} \footnote{In such a case one could introduce the notion of $(\kappa,\tilde{\kappa},\rho)$-Yang models, where $\rho$ is Born self-dual parameter.}.

The plan of our paper is the following.

In Sect. 2 we recall algebraic descriptions of Snyder and Yang models, and show how by using the generalized Born map one can derive from $\kappa$-Snyder model the ($\kappa,\tilde{\kappa}$)-Yang model.

In Sect. 3 we introduce the algebra of ($\kappa,\tilde{\kappa}$)-Yang model as covariant under the generalized Born map (\ref{bm1}). By extending the method presented in \cite{MM_PLB814}-\cite{LMMP_PLB838} we show that the relations describing ($\kappa,\tilde{\kappa}$)-Yang model can be obtained consistently by the suitable linear transformation of the generators of $\hat{o}(1,5)$. We point out that both  $D=4$ standard Yang model and the new $D=4$ $(\kappa,\tilde{\kappa})$-Yang model can be described by two different realizations of $\hat o(1,5)$ algebra linked by the linear map.

In the last Sect. 4 we include brief conclusions and we present two new ideas which we believe should be further developed.

\section{From $\kappa$-Snyder to doubly $\kappa$-dependent Yang models}

The NC algebraic structures in $D=4$ Snyder and Yang type models are described  by Lie algebras $\hat{o}(1,4)$ ($D=4$ dS algebra) and $\hat{o}(1,5)$ ($D=5$ dS algebra which is isomorphic to $D=4$ Euclidean conformal algebra).
In Snyder model, by using $D=4$ dS algebra generators $\hat{M}_{ab}=(\hat{M}_{\mu\nu},\hat{M}_{4\mu})$ ($a,b=0,1,\ldots,4$), one postulates the following identification of NC space-time coordinates ($\mu,\nu=0,1,2,3$):
\begin{equation}
\hat{M}_{4\mu }=M \hat{x}_\mu\label{x_id}
\end{equation}
where $M$ denotes the inverse of the elementary length parameter which plays the role of dimensionful mass-like deformation parameter, frequently identified with the Planck mass. The following set of algebraic relations describes the $D=4$ Snyder model \cite{Snyder}\footnote{Following the original formulation in ref.~\cite{Snyder, Yang} of Snyder and Yang models,  we will expose explicitly the dependence of algebraic formulas on the Planck constant $\hbar$ (see also \cite{Brodsky, LMMP_2212}). Such $\hbar$-dependent algebras provide quantum-mechanical formulation of Snyder and Yang models.} with $\eta_{\mu\nu}=diag(-1,1,1,1)$:
\begin{equation}
\lbrack \hat{x}_{\mu },\hat{x}_{\nu }]=\frac{i\hbar}{M^2}\hat{M}_{\mu
\nu },  \label{snyderx}
\end{equation}
\begin{equation}
\lbrack \hat{M}_{\mu \nu },\hat{x}_{\rho }]=i\hbar (\eta _{\mu \rho }\hat{x}%
_{\nu }-\eta _{\nu \rho }\hat{x}_{\mu }), \label{snyderMx}
\end{equation}
\begin{equation}
\lbrack \hat{M}_{\mu \nu },\hat{M}_{\rho \tau }]=i\hbar (\eta _{\mu \rho
}\hat{M}_{\nu \tau }-\eta _{\mu \tau }\hat{M}_{\nu \rho }+\eta _{\nu \tau }\hat{M}_{\mu \rho
}-\eta _{\nu \rho }\hat{M}_{\mu \tau }) \label{snyderMM}
\end{equation}
where relations (\ref{snyderMx}) express the Lorentz covariance of Snyder model and (\ref{snyderMM}) describes the relativistic Lorentz extension of quantum mechanical nonrelativistic angular momentum $\hat o(3)$ algebra, with $\hbar$-dependence used as in standard books on Quantum Mechanics (see e.g. \cite{Landau,Weinberg}).

The Yang model is obtained if we supplement the relations (\ref{snyderx}-\ref{snyderMM}) by the algebraic relations for quantum relativistic fourmomenta $\hat{q}_\mu$:
\begin{equation}  \label{qq}
\lbrack \hat{q}_{\mu },\hat{q}_{\nu }]={\frac{i\hbar }{R^{2}}}\hat{M}_{\mu
\nu },\end{equation}
\begin{equation}
\lbrack \hat{M}_{\mu\nu},\hat{q}_{\rho}]=i\hbar (\eta _{\mu\rho}\hat{q}_{\nu}-\eta
_{\nu\rho}\hat{q}_{\mu})\label{Mq}
\end{equation}
where in astrophysical applications $R$ describes the cosmological $D=4$ dS radius. From the relations (\ref{x_id}-\ref{Mq}) it follows that we obtain $\hat{o}(1,5)$ Lie algebra with $\hat{M}_{AB}=(\hat{M}_{\mu\nu},\hat{M}_{4\mu},\hat{M}_{5\mu}, \hat{M}_{45})$ ($A,B=0,1,\ldots 5$) if we assign the generators $\hat{M}_{5\mu}$, $ \hat{M}_{45}$ to the quantum fourmomenta variables $\hat{q}_\mu$ as follows, see e.g. \cite{LMMP_2212}:
\begin{equation}\label{q_id}
\hat{M}_{5\mu }=R\hat{q}_\mu,\qquad \hat{M}_{45}=MR\hat{r}
\end{equation}
where in $D=4$ the rescaled generator $\hat{r}$ describes the quantum internal $\hat{o}(2)$ symmetry acting on the $\hat o(2)$  doublet representation $(\hat{x}_\mu,\hat{q}_\mu)$:
\begin{equation}\label{Ixq}
[\hat{r},\hat{x}_\mu]=\frac{i\hbar}{M^2} \hat{q}_\mu, \qquad [\hat{r},\hat{q}_\mu]=-
\frac{i\hbar}{R^2} \hat{x}_\mu.
\end{equation}
The dimensionful factors $R^2$ and $M^2$ imply that the generator $\hat{r}$ is dimensionless, and determines the following basic relativistic Heisenberg algebra relation:
\begin{equation}\label{xqI}
[\hat{x}_\mu, \hat{q}_\nu]=i\hbar\eta_{\mu\nu}\hat{r}
\end{equation}
where the case with
$\hat{r}=1$ corresponds to the canonical commutation relations.

One should observe that the Yang model algebra described by the relations (\ref{snyderx}-\ref{Mq}) and
(\ref{Ixq}-\ref{xqI}) is covariant under the following Born map $B$ \cite{Born1,Born2,Freidel}
{
\begin{equation}\label{born}
B:\quad \hat{x}_\mu\rightarrow \hat{q}_\mu,\quad \hat{q}_\mu\rightarrow - \hat{x}_\mu,\quad M\leftrightarrow {R},\quad\hat{M}_{\mu\nu}\leftrightarrow \hat{M}_{\mu\nu},\quad \hat{r}\leftrightarrow\hat{r}
\end{equation}}
where $B$ is a pseudo-involution satisfying the relation $B^4=1$, which permits to define the Yang model as the Born-map extension of the Snyder model.

$\kappa$-Snyder model, proposed in \cite{kappa-Snyder1,kappa-Snyder}, has the following two-parameter extension of the relations (\ref{snyderx}-\ref{snyderMx}): 
\begin{equation}
\lbrack \hat{x}_{\mu },\hat{x}_{\nu }]=i\hbar\left[\frac{1}{M^2}\hat{M}_{\mu
\nu}+\frac{1}{\kappa}(a_\mu \hat{x}_\nu - a_\nu \hat{x}_\mu)\right],  \label{snyderkx}
\end{equation}
\begin{equation}
\lbrack \hat{M}_{\mu \nu },\hat{x}_{\rho }]=i\hbar \left[\eta _{\mu \rho }\hat{x}_{\nu }-\eta _{\nu \rho }\hat{x}_{\mu }
+\frac{1}{\kappa}(a_\mu \hat{M}_{\rho\nu} - a_\nu \hat{M}_{\rho\mu})\right].  \label{snyderMkx}
\end{equation}
Recently various properties of this model were investigated in \cite{MM_PLB814,MM_PRD104, MP_2021,LMMP_PLB838}.
The constant dimensionless four-vector $a_\mu$ permits to select three types of the $\kappa$-deformations of quantum Minkowski spaces: time-like (or standard one) if $a_\mu a^\mu=-1$, tachyonic if $a_\mu a^\mu =1$ and light-like if $a_\mu a^\mu=0$, corresponding to the metric signature we have chosen. If we put $M\to\infty$ in (\ref{snyderkx}) we obtain the generalized $a_\mu$-dependent $\kappa$-deformed Minkowski space-time, with $\hat{x}=a^\mu \hat{x}_\mu$
describing the unique NC quantum coordinate.

The main aim of this paper is to introduce the new doubly $\kappa$-dependent Yang model, obtained from the $\kappa$-Snyder model by adding the $\kappa$-Minkowski type terms to quantum fourmomenta sector.
We add them by postulating that:
\begin{equation}\label{qqk}
[\hat{q}_\mu,\hat{q}_\nu]=i\hbar \left[\frac{\hat{M}_{\mu\nu}}{R^2}+\tilde{\kappa}(b_\mu\hat{q}_\nu-b_\nu\hat{q}_\mu)\right].
\end{equation}
The relation (\ref{qqk}) can be obtained from relation (\ref{snyderkx}) by the use of the generalized Born map $\tilde B$ which one gets by adding to (\ref{born}) the following relations \footnote{{We obtain as a special case the generalized $\kappa$-dependent Yang model if we put in (\ref{qqk}-\ref{born2}) $\kappa=\tilde{\kappa}$ and $a_\mu=b_\mu$. Further, one can as well postulate that $\kappa=M$.}}
\begin{equation}\label{born2}
\tilde B:\quad a_\mu\rightarrow b_\mu, \quad b_\mu\rightarrow - a_\mu,\quad \kappa\leftrightarrow \frac{1}{\tilde{\kappa}}.
\end{equation}
Further, after using the generalized Born map $\tilde B$ given by the relations (\ref{born}), (\ref{born2}), one gets from (\ref{snyderMkx}) the following $\kappa$-dependence of covariance relations (\ref{Mq}) for quantum fourmomenta
\begin{equation}
[ \hat{M}_{\mu\nu},\hat{q}_{\rho}]=i\hbar \left[\eta _{\mu\rho}\hat{q}_{\nu}-\eta
_{\nu\rho}\hat{q}_{\mu}+\tilde{\kappa}(b_\mu\hat{M}_{\rho\nu}-b_\nu\hat{M_{\rho\mu}})\right].\label{Mqkappa}
\end{equation}
Finally, it follows that the relations (\ref{Ixq}), (\ref{xqI}), due to their selfduality under the map (\ref{born}), remain the same in $(\kappa,\tilde{\kappa})$-Yang model.
\medskip

\section{Doubly $\kappa$-dependent Yang model and new quantum relativistic phase spaces}
It is already known since 1947 (see \cite{Yang}) that the algebra describing $D=4$ Yang model is spanned by the $\hat{o}(1,5)$ Lie algebra generators
\begin{equation}
\lbrack \hat{M}_{AB},\hat{M}_{CD}]=i\hbar (\eta _{AC}\hat{M}_{BD}-\eta _{AD}\hat{M}_{BC}+\eta _{BD}\hat{M}_{AC}-\eta _{BC }\hat{M}_{AD}) \label{YangMM}
\end{equation}
where $\eta_{AB}=diag(-1,1,\ldots,1),\quad (A,B=0,1,\ldots, 5)$.
We will show that by the generalization of the recent description of $\kappa$-Snyder model \cite{MM_PLB814}-\cite{LMMP_PLB838} one can describe $(\kappa,\tilde{\kappa})$-Yang models as the Lie algebra $\hat{o}(6;g_{AB})$:
 \begin{equation}
\lbrack \hat{M}^{(Y)}_{AB},\hat{M}^{(Y)}_{CD}]=i\hbar (g^{(Y)}_{AC}\hat{M}^{(Y)}_{BD}-g^{(Y)}_{AD}\hat{M}^{(Y)}_{BC}+g^{(Y)}_{BD}\hat{M}^{(Y)}_{AC}-g^{(Y)}_{BC }\hat{M}^{(Y)}_{AD}) \label{Yang_gMM}
\end{equation}
where the symmetric metric components $g^{(Y)}_{AB}$ with the signature $(-1,1,\ldots,1)$ depend on five deformation parameters $(M,R,\kappa,\tilde{\kappa},\rho)$  $([M]=L^{-1},[R]=L,[\kappa]=[\tilde{\kappa}]=L^{-1},[\rho]=L^0)$ and a pair of constant dimensionless four-vectors $a_A=(a_\mu,0,0), b_A=(b_\mu,0,0)$, $\mu=0,1,2,3$ which respectively determine the type of $\kappa$-dependence in quantum $D=4$ space-time and $D=4$ quantum  fourmomenta sectors of Yang algebra. The metric $g^{(Y)}_{AB}$ is determined by the following assignments of the generators (see also (\ref{x_id}) and (\ref{q_id})):
\begin{equation}
M^{(Y)}_{AB}=\left(\hat{M}_{\mu\nu},\ \hat{M}^{(Y)}_{4\mu}=M\hat{x}_\mu,\ \hat{M}^{(Y)}_{5\mu}=R\hat{q}_\mu,\ \hat{M}^{(Y)}_{45}=MR\hat{r}\ \right)\label{M_Y_AB}
\end{equation}
where $[M^{(Y)}_{AB}]=L^0$ (dimensionless), in consistency with relation (\ref{Yang_gMM}), with $\hat{M}_{\mu\nu}$ describing $D=4$ Lorentz algebra and the scalar $\hat{r}$ providing the generator of the $\hat{o}(2)$ internal symmetries.
Relations (\ref{Yang_gMM}) are describing the $(\kappa,\tilde{\kappa})$-Yang model if we insert the following components of the $D=6$ metric tensor:
\begin{equation}\label{g_matrix}
g_{AB}^{\left( Y\right) }=\left(
\begin{array}{ccc}
\eta _{\mu \nu } & g_{\mu 4}^{\left( Y\right) } & g_{\mu 5}^{\left( Y\right)
} \\
g_{4\nu }^{\left( Y\right) } & g_{44}^{\left( Y\right) } & g_{45}^{\left(
Y\right) } \\
g_{5\nu }^{\left( Y\right) } & g_{54}^{\left( Y\right) } & g_{55}^{\left(
Y\right) }%
\end{array}%
\right)
\end{equation}
where\footnote{One can choose three types of constant four-vectors $a_{\mu
}$ and $b_{\mu }$, with Lorentz-invariant lengths $\left( -1,0,1\right) $,
which select three types of $\kappa $ and $\tilde{\kappa}$-Minkowski terms: time-like
(or standard one), light-like and tachyonic. The
four-vectors $a_{\mu }$ and $b_{\mu }$ determine the quantum NC phase space
components $a^{\mu }\hat{x}_{\mu }$ and $b^{\mu }\hat{p}_{\mu }$, which due
to the double $\kappa -$dependence $\left( \kappa \neq 0,\tilde{\kappa}%
\neq 0\right) $ break explicitly the Lorentz covariance (compare with (\ref{snyderMkx}) and (\ref{Mqkappa})).}
\begin{equation}\label{g_coeff}
g_{\mu 4}^{\left( Y\right) }=g_{4\mu }^{\left( Y\right) }=\frac{M}{%
\kappa }a_\mu,\,\quad g_{\mu
5}^{\left( Y\right) }=g_{5\mu }^{\left( Y\right) }=R\tilde{\kappa}b_{\mu
},\quad g_{45}^{\left( Y\right) }=g_{54}^{\left( Y\right) }=\rho ,\quad
g_{44}^{\left( Y\right) }=g_{55}^{\left( Y\right) }=1
\end{equation}
with the pair of length parameters ($\lambda =M^{-1},R$) (or equivalently the pair of mass parameters $(M=\lambda^{-1}, \tilde M=R^{-1})$),
 the mass-like parameters $\left( \kappa ,\tilde{\kappa}\right) $ and the dimensionless parameter $\rho $, i.e $g^{(Y)}_{AB}$ are dimensionless ($[g^{(Y)}_{AB}]=L^0$) in consistency with relations (\ref{Yang_gMM}).

The algebra (\ref{Yang_gMM}) for any choice of symmetric metric $g^{(Y)}_{AB}$ satisfies two
important properties:
\smallskip

i) By direct calculation one can show that the Lie algebra (\ref{Yang_gMM}) satisfies
Jacobi identities.

ii) For any nondegenerate symmetric metric $g^{(Y)}_{AB}$ with the signature described
by diagonal matrix $\eta _{AB}$ one can find $\left( 6\times 6\right)$-dimensional linear map $\mathbb{S}=S_{AB}$ satisfying the relation
\begin{equation}\label{gtrans}
\mathbf{g}^{\left( Y\right) }=\mathbb{S}\mathbb{\eta} \mathbb{S}^{T},\qquad \mathbf{g}^{\left( Y\right) }\equiv
\mathbf{g}_{AB}^{\left( Y\right) },\qquad \mathbb{\eta}=\eta_{AB}.
\end{equation}
One can also relate the Lie algebras (\ref{YangMM}) and (\ref{Yang_gMM}) by the following maps
\begin{equation}\label{Mtrans}
\hat{M}^{(Y)}_{AB}=(\mathbb{S}\,\hat{\mathbb{M}}^{(0)}\mathbb{S}^T)_{AB} \qquad\longleftrightarrow\qquad \hat{M}^{(0)}_{AB}=(\mathbb{S}^{-1}\hat{\mathbb{M}}^{(Y)}(\mathbb{S}^T)^{-1})_{AB}.
\end{equation}
We point out that the map \eqref{Mtrans} describes a simple linear transformation between realizations of D=6 dimensional algebra $\hat o(1,5)$. {In $D=4$ physical dimensions}, it describes the passage from the standard D=4 Yang model
(i.e. algebra \eqref{YangMM}) to new D=4 doubly-$\kappa$-dependent Yang model, where coordinates and momenta do not commute and their commutator is proportional to the sum of Lorentz generators and $\kappa$-Minkowski terms in space-time sector (and respectively $\tilde{\kappa}$-Minkowski type terms in fourmomenta sector), i.e. the algebraic relations \eqref{Yang_gMM}-\eqref{g_coeff}.

Additionally, we observe that the matrix $\mathbb{S}$ satisfying relations (\ref{gtrans}), (\ref{Mtrans}) is not unique, with arbitrariness described by the pseudoorthogonal matrix $\mathbb{O}$, where $\mathbb{O}\eta\mathbb{O}^T=\eta.$ For concrete choice (\ref{gtrans}) of the matrix $\mathbf{g}^{(Y)}$ we choose 
$6\times 6$ matrix $\mathbb{S}$ parametrized as follows\footnote{For the simplicity of formulae in (\ref{esse}), we introduce the shorthand notation
$$
g_{\mu 4}^{\left( Y\right) }=g_{4\mu }^{\left( Y\right) }=g_{\mu },\quad
g_{\mu 5}^{\left( Y\right) }=g_{5\mu }^{\left( Y\right) }=h_{\mu },
$$
$$
g_{44}^{\left( Y\right) }=g_{4},\quad g_{45}^{\left( Y\right)
}=g_{54}^{\left( Y\right) }=\rho ,\quad g_{55}^{\left( Y\right) }=h_{5}.
$$
}
\begin{equation}\label{esse}
\mathbb{S}=\left(
\begin{array}{cccccc}
-1 & 0 & 0 & 0 & 0 & 0 \\
0 & 1 & 0 & 0 & 0 & 0 \\
0 & 0 & 1 & 0 & 0 & 0 \\
0 & 0 & 0 & 1 & 0 & 0 \\
g_{0} & g_{1} & g_{2} & g_{3} & a & d \\
h_{0} & h_{1} & h_{2} & h_{3} & c & b%
\end{array}%
\right) \end{equation}
with parameters $a$, $b$, $c$, $d$ satisfying the conditions
\begin{eqnarray}\label{25a}
&&a^2+d^2=g_4-g_\mu g^\mu=A,\cr
&&b^2+c^2=h_5-h_\mu h^\mu=B,\cr
&&ac+bd=\rho-g_\mu h^\mu=C.
\end{eqnarray}
We can pass to lower triangular $\mathbb{S}$ matrix if we put $d=0$ in the formulae (\ref{esse},\ref{25a}). In such a case 
the set of equations (\ref{25a}) has the following solutions:
\begin{equation}
a=\epsilon\sqrt{A},\qquad b=\epsilon'\sqrt{B-\frac{C^2}{A}},\qquad c=\epsilon\frac{C}{\sqrt{A}},
\end{equation}
with $\epsilon,\epsilon'=\pm1$.

In $(\kappa,\tilde{\kappa})$-Yang models one can select nine classes of double $\kappa$-dependence based on the Lorentz-covariant normalized length values of the dimensionless fourvectors $a_\mu, b_\mu$ which are related with the fourvectors $g_\mu$ and $h_\mu$ from the matrix (\ref{esse}) if we choose
\begin{equation}
g_\mu=\frac{M}{\kappa}a_\mu \rightarrow A=1-\frac{M^2}{\kappa^2}a_\mu a^\mu,
\end{equation}
\begin{equation}
{h}_\mu=R\tilde{\kappa}b_\mu \rightarrow B=1-R^2\tilde{\kappa^2}b_\mu b^\mu
\end{equation}
and 
\begin{equation}
C=\rho-\frac{M}{\kappa}R\tilde{\kappa}a_\mu b^\mu.
\end{equation}
The nine classes of double $\kappa$-dependence are obtained by the choices of the parameters $ {\epsilon},  {\epsilon}'=(\pm 1,0)$, where $a_\mu a^\mu = {\epsilon}$ and
$b_\mu b^\mu = {\epsilon}'$.

The missing  algebraic relations of $(\kappa,\tilde{\kappa})$-Yang model, which describe the modified $D=4$ Heisenberg algebra sector of $\hat{o}(6;g^{(Y)}_{AB})$ by the generalization of relations (\ref{Ixq}), (\ref{xqI}) look as follows:
\begin{equation}\label{xqd}
[\hat{x}_\mu,\hat{q}_\nu]=i\hbar\left(\eta_{\mu\nu}\hat{r}+\tilde{\kappa}b_\mu\hat{x}_\nu -\frac{a_\nu}{\kappa} \hat{q}_\mu +\frac{\rho}{MR}\hat{ M}_{\mu\nu}\right),
\end{equation}
\begin{equation}\label{Ixd}
[\hat{r},\hat{x}_\mu]=i\hbar\left(\frac{1}{M^2}\hat{q}_\mu-\frac{1}{MR}\rho\hat{x}_\mu - \frac{a_\mu}{\kappa}\hat{r}\right),
\end{equation}
\begin{equation}\label{Iqd}
[\hat{r},\hat{q}_\mu]=i\hbar\left(-\frac{1}{R^2}\hat{x}_\mu+\frac{1}{MR}\rho\hat{q}_\mu - \tilde{\kappa} b_\mu\hat{r}\right).
\end{equation}
Additionally, we have
\begin{equation}
\lbrack \hat{r}, \hat{M}_{\mu \nu }]=-i\hbar \left[\frac{1}{\kappa}(a_\mu\hat{q}_{\nu }-
a_\nu\hat{q}_\mu)-\tilde{\kappa}(b_\mu
\hat{x}_{\nu } - b_\nu \hat{x}_{\mu})\right],  \label{snyderBD_IM}
\end{equation}
i.e. we see that the  internal  and Lorentzian generators do not commute with each other.

It can be checked that the relations (\ref{xqd})-(\ref{snyderBD_IM}) are self dual under the generalized Born map (\ref{born}), extended by (\ref{born2}). Moreover, it can be shown that the relations (\ref{xqd})-(\ref{snyderBD_IM}) can be derived as the general solutions of the Jacobi identities for generators $\hat{x}_\mu$, $\hat{q}_\mu$, $\hat{r}$ and $\hat{M}_{\mu\nu}$.

If in the equations (\ref{snyderkx})-(\ref{qqk}), (\ref{Mqkappa}) and (\ref{xqd})-(\ref{Iqd}) we perform the limits $R\to\infty$, $\tilde{\kappa}\to 0$, $\rho\to 0$ we get standard two-parameter $\kappa$-Snyder model, describing NC quantum space-time with $\hat x_\mu$ \cite{MM_PLB814}-\cite{LMMP_PLB838}. However, by taking the limits $M\to\infty, \kappa\to\infty$ and $\rho\to 0$, we obtain the Born-dual $\tilde{\kappa}$-Snyder model in quantum fourmomenta space $\hat{q}_\mu$, with the new type of $\kappa$-dependence (we call it $\tilde{\kappa}$-dependence;  see also  (\ref{qqk}),(\ref{Mqkappa})).

The relations (\ref{snyderkx}), (\ref{snyderMkx}) define the NC $\kappa$-Snyder quantum space-time which is described by the subalgebra $\hat{o}(5;g^{(Y)}_{ab})$, $(a,b=0,1,\ldots,4)$ of the Yang algebra $\hat{o}(6;g^{(Y)}_{AB})$, $(A=(a,5), B=(b,5)$ (see (17)).
Similarly, the Born-dual subalgebra $\hat{o}(5;g^{(Y)}_{\tilde{a}\tilde{b}})$ 
of Yang algebra (17), where $(\tilde{a},\tilde{b}=0,1,2,3,5)$, describes the NC $\tilde{\kappa}$-Snyder relations in quantum fourmomenta sector (see relations (\ref{qqk}),(\ref{Mqkappa})).

\section{Outlook}

In this paper we have proposed the new relativistic quantum phase spaces by introducing $\kappa$-extensions of the standard Yang model
which define doubly $\kappa$-dependent Yang models (i.e. $(\kappa,\tilde{\kappa})$-Yang models). In such  models there appear additional $\kappa$-Minkowski terms, linked by the generalized Born map (\ref{born}), (\ref{born2}), which provide the standard $\kappa$-Minkowski type terms in commutativity relations between quantum relativistic space-time coordinates $\hat{x}_\mu$ (see (\ref{snyderkx}), (\ref{snyderMkx})) and introduce the new $\tilde{\kappa}$-Minkowski type terms in quantum  fourmomenta commutation relations (see (\ref{qqk}), (\ref{Mqkappa})).

The doubly $\kappa$-dependent Yang model, proposed in this paper, is described by the following five parameters ($M,R,\kappa,\tilde{\kappa},\rho$):
\begin{itemize}
\item[-]  mass-like parameter $M$ ($[M]=L^{-1}$) describing the constant curvature in quantum space-time sector (e.g. $M$ can be identified with the Planck mass)
\item[-] parameter $R$ ($[R]=L^{1}$) defining the constant curvature in quantum fourmomenta sector (e.g. $R$ can be linked to the radius of the Universe)
\item[-] parameters $\kappa$ and $\tilde{\kappa}$ which describe two independent modifications of quantum space-time and quantum fourmomenta sectors, respectively
\item[-] the fifth dimensionless parameter $\rho$ ($[\rho]=L^{0}$) parametrising, in the commutator $[\hat{x}_\mu,\hat{q}_\nu]$, the term proportional to $M_{\mu\nu}$ \footnote{We observe that
such term where $\rho=1$ occurs in the so-called Triple Special Relativity (TSR) model \cite{TSR}, which however, contrary to the case of Yang model, cannot be reformulated algebraically as described by the particular realizations of any classical Lie algebra (see \cite{MM22_PLB}, \cite{MM22_JMP}).}.
\end{itemize}
We have shown in Sect. 3 that the algebraic structure of doubly $\kappa$-dependent Yang model can be derived from the $\hat{o}(1,5)$
algebra if we select the suitable realizations of its generators by proper choice of the matrix $\mathbb{S}$ (see (\ref{gtrans}- \ref{esse})). 
 
Finally, we propose two ways which could lead to the valuable generalizations of Snyder and Yang models.

\begin{itemize}
\item[-] Firstly, we propose the generalization of Yang models to the Kaluza-Klein geometries in $D=(1,3+2N)$ with Lorentzian signature and $\hat{o}(2N)$ internal symmetries (in particular if $N=1$ we obtain the standard Yang model with Born-extended internal Abelian $\hat{o}(2)$ symmetry). Such models should be useful in studies of the new unification models of gravity and particle physics which employ the higher-dimensional Lorentz algebras $\hat{o}(1,3+2N)$ (e.g. for $N=5$ see \cite{cham}, for $N=7$ see \cite{zou24}).
\item[-]  Our second idea for the future is to consider the quantum-deformed Hopf algebras $\hat{o}(1,4)$ and $\hat{o}(1,5)$ as defining algebraically new quantum Snyder and Yang models. In particular, it will be interesting to consider particular deformations of $\hat{o}(1,4)$ and $\hat{o}(1,5)$ in which the Lorentz subalgebra $\hat{o}(1,3)$ is deformed, but still remains the Hopf subalgebra of quantum $\hat{o}(1,4)$ and $\hat{o}(1,5)$ algebras (see e.g. \cite{Angel}).
\end{itemize}

\section*{Acknowledgements}
J. Lukierski, A. Pacho{\l} and M. Woronowicz acknowledge the support of the Polish NCN grant 2022/45/B/ST2/01067.
S. Mignemi recognizes the support
of Gruppo Nazionale di Fisica Matematica.  A. Pacho{\l} acknowledges COST Action CaLISTA CA21109.


\begin{thebibliography}{99}
\bibitem{DFR} 
S.~Doplicher, K.~Fredenhagen and J.~E.~Roberts,
Commun. Math. Phys. \textbf{172} (1995), 187-220
[arXiv:0303037 [hep-th]].
\bibitem{Majid2009} 
S.~Majid,
book chapter in D. Oriti (ed.) "Approaches to Quantum Gravity", Cambridge Univ. Press (2009), p.466
[arXiv:0604130 [hep-th]].
\bibitem{Low}
S.~G.~Low,
J. Phys. A \textbf{35} (2002), 5711-5730
[arXiv:0101024 [math-ph]].


\bibitem{Lukierski2021PoS}
J.~Lukierski and M.~Woronowicz,
PoS {CORFU2021} (2022), 290
[arXiv:2204.07787 [hep-th]].

\bibitem{AP_EPJC2023}
A.~Pacho\l{} and A.~Wojnar, Eur. Phys. J. C 83, 1097 (2023)

\bibitem{Snyder} H.~S.~Snyder, 
Phys. Rev. \textbf{71}, 38-41 (1947) 
\bibitem{Yang} C.N. Yang, Phys. Rev. D \textbf{47} (1947), 874
\bibitem{Bronstein}
M. Bronstein, JETP \textbf{9}, 140152 (1936)

\bibitem{Chamseddine}
A.~H.~Chamseddine and A.~Connes,
Phys. Rev. Lett. \textbf{77} (1996), 4868-4871
[arXiv:9606056 [hep-th]].

\bibitem{Aschieri}
P.~Aschieri, M.~Dimitrijevic, F.~Meyer and J.~Wess,
Class. Quant. Grav. \textbf{23} (2006), 1883-1912
[arXiv:0510059 [hep-th]].



\bibitem{Drinfeld}
V.~G.~Drinfeld,
Contribution to:
    1986 International Congress of Mathematics, Berkley Press, vol. 1, 798-820 (1986)
\bibitem{Connes}
A. Connes, "Noncommutative geometry", Acad. Press (1994)
\bibitem{Beggs}
E. J. Beggs, S. Majid, "Quantum Riemannian Geometry", Grundlehren der mathematischen Wissenschaften 355, Springer International Publishing (2020)
\bibitem{Arzano}
M.~Arzano and J.~Kowalski-Glikman,
``Deformations of Spacetime Symmetries: Gravity, Group-Valued Momenta, and Non-Commutative Fields,''
2021,
ISBN 978-3-662-63095-2, 978-3-662-63097-6



\bibitem{17a}
S.~Mignemi,
Class. Quant. Grav. \textbf{26} (2009), 245020
\bibitem{17b}
R.~Banerjee, K.~Kumar and D.~Roychowdhury,
JHEP \textbf{03} (2011), 060
[arXiv:1101.2021 [hep-th]].

\bibitem{5a} 
M.~V.~Battisti and S.~Meljanac,
Phys. Rev. D \textbf{79} (2009), 067505
[arXiv:0812.3755 [hep-th]].
\bibitem{5aa}
M.~V.~Battisti and S.~Meljanac,
Phys. Rev. D \textbf{82} (2010), 024028
[arXiv:1003.2108 [hep-th]].


\bibitem{Born1}
M. Born, Proceedings of the Royal Society (London) \textbf{A165}, 291 (1938)
\bibitem{Born2}
M. Born, Rev. Mod. Phys. 21, 463 (1949)


\bibitem{1991pn}
    J. Lukierski, H. Ruegg, A. Nowicki and Valerij N. Tolstoy,
    Phys. Lett. B \textbf{264} 331 (1991) 
\bibitem{MRkappa}
	S. Majid, H. Ruegg, Phys. Lett. B \textbf{334}, 348 (1994),
 	[arXiv: hep-th/9405107 [hep-th]].
\bibitem{LRkappa}
	J. Lukierski, H. Ruegg, Phys. Lett. B \textbf{329}, 189 (1994),
 	[arXiv:  hep-th/9310117].
\bibitem{LRZkappa}
	J. Lukierski, H. Ruegg, W.J. Zakrzewski, Ann. Phys. \textbf{243}, 90 (1995),
 	[arXiv:  hep-th/9312153].

\bibitem{Planckmass} 
J.~Kowalski-Glikman,
Phys. Lett. A \textbf{286} (2001), 391-394
[arXiv:hep-th/0102098 [hep-th]].

\bibitem{kappa-Snyder1}
S.~Meljanac, D.~Meljanac, A.~Samsarov and M.~Stojic,
[arXiv:0909.1706 [math-ph]].

\bibitem{kappa-Snyder}
S.~Meljanac, D.~Meljanac, A.~Samsarov and M.~Stojic,
Mod. Phys. Lett. A \textbf{25} (2010), 579-590
[arXiv:0912.5087 [hep-th]].


\bibitem{Melj2011}
S.~Meljanac, D.~Meljanac, A.~Samsarov and M.~Stojic,
Phys. Rev. D \textbf{83} (2011), 065009
[arXiv:1102.1655].

\bibitem{BP_EPJC}
A.~Borowiec and A.~Pacho\l,
Eur. Phys. J. C \textbf{74} (2014) no.3, 2812
[arXiv:1311.4499].

\bibitem{BP_SIGMA2014}
A.~Borowiec and A.~Pacho\l,
SIGMA \textbf{10} (2014), 107
[arXiv:1404.2916].




\bibitem{MM_PLB814} S.~Meljanac and S.~Mignemi,
Phys. Lett. B \textbf{814}, 136117 (2021)
[arXiv:2101.05275 [hep-th]].

\bibitem{MM_PRD104} S.~Meljanac and S.~Mignemi,
Phys. Rev. D \textbf{104}, no.8, 086006 (2021)
[arXiv:2106.08131 [physics.gen-ph]].


\bibitem{MP_2021} S.~Meljanac and A.~Pacho\l {},
Symmetry \textbf{13}, no.6, 1055 (2021) 
[arXiv:2101.02512 [hep-th]].





\bibitem{LMMP_PLB838}
J.~Lukierski, S.~Meljanac, S.~Mignemi and A.~Pacho\l,
Phys. Lett. B \textbf{838} (2023), 137709
[arXiv:2208.06712].


\bibitem{lukpa}
J. Lukierski, A. Pacho\l, PoS CORFU 2023.

\bibitem{MMM2024}
T. Martinić-Bilać, S. Meljanac, S. Mignemi, [arXiv: 2404.01792 [hep-th]].

\bibitem{TSR}
J.~Kowalski-Glikman and L.~Smolin, 
Phys. Rev. D \textbf{70} (2004), 065020 
[arXiv:0406276 [hep-th]].



\bibitem{Brodsky}
S.~J.~Brodsky and P.~Hoyer,
Phys. Rev. D \textbf{83} (2011), 045026
[arXiv:1009.2313 [hep-ph]].

\bibitem{LMMP_2212} J.~Lukierski, S.~Meljanac, S.~Mignemi and
A.~Pacho\l,
Phys. Lett. B \textbf{847} (2023), 138261
[arXiv:2212.02316 [hep-th]].

\bibitem{Landau}
L.D. Landau, E.M. Lifshitz, "Quantum Mechanics (Nonrelativistic theory)", 2nd edition, Pergamon Press (1965)
\bibitem{Weinberg}
S. Weinberg, "Lectures on Quantum Mechanics", 2nd edition, Cambridge University Press (2019)






\bibitem{Freidel}
L.~Freidel, R.~G.~Leigh and D.~Minic,
Phys. Lett. B \textbf{730} (2014), 302-306

[arXiv:1307.7080 [hep-th]].




\bibitem{MM22_PLB}
S.~Meljanac and S.~Mignemi,
Phys. Lett. B \textbf{833} (2022), 137289
[arXiv:2206.04772 [hep-th]].

\bibitem{MM22_JMP}
S.~Meljanac and S.~Mignemi,
J. Math. Phys. \textbf{64} (2023) no.2, 023505
[arXiv:2211.11755 [gr-qc]].

\bibitem{cham}
A.H. Chamseddine, V. Mukhanov, JHEP \textbf{03} (2016) 020, [arXiv: 1602.02295 [hep-th]].
\bibitem{zou24}
 D. Roumelioti, S. Stefas, G. Zoupanos,  [arXiv: 2403.17511 [hep-th]].

\bibitem{Angel}
A.~Ballesteros, I.~Gutierrez-Sagredo and F.~J.~Herranz,
Class. Quant. Grav. \textbf{39} (2022) no.1, 015018
[arXiv:2108.02683 [math-ph]].

\end{thebibliography}
\end{document}